\documentclass[aps,prd,fleqn,superscriptaddress]{revtex4}
\usepackage{graphicx,xcolor,natbib,braket,float}
\usepackage{amsmath,amssymb,amsfonts,cases}
\newcommand{\bse}{\begin{subequations}}
\newcommand{\ese}{\end{subequations}}
\newcommand{\be}{\begin{equation}}
\newcommand{\ee}{\end{equation}}
\newcommand{\bea}{\begin{eqnarray}}
\newcommand{\eea}{\end{eqnarray}}
\newcommand{\ba}{\begin{array}}
\newcommand{\ea}{\end{array}}

\usepackage[colorlinks=true, linkcolor=blue, bookmarks=true]{hyperref}

\begin{document}
\title{Holographic Entanglement and Emergent Gravity}
\author{Mohammad Ali-Akbari\footnote{$\rm{m}_{-}$aliakbari@sbu.ac.ir}}
\affiliation{Department of Physics, Shahid Beheshti University, 1983969411, Tehran, Iran}

\begin{abstract}
We investigate the emergence of bulk geometry from boundary entanglement entropy in the context of holographic duality. Considering a thermal $1+1$-dimensional conformal field theory dual to the BTZ black hole, we compute four distinct holographic entanglement measures: the standard spacelike HEE, its complementary spacelike counterpart HEE$^s$, the timelike HTEE, and its complementary timelike counterpart HEE$^t$. These measures probe both the exterior and interior regions of the black hole horizon. Remarkably, by taking appropriate derivatives of these entanglement entropies with respect to the boundary intervals and the turning point of the extremal surface, we reconstruct the full BTZ metric, including the radial component $g_{zz}$, without imposing the Einstein equations. The reconstruction yields the correct metric components in all regions of the geometry. We further show that in the zero-temperature limit, our results consistently reduce to the pure AdS$_3$ metric and the familiar vacuum CFT$_2$ entanglement entropies. Our findings demonstrate that entanglement entropy data alone contains sufficient information to uniquely determine the dual gravitational background, offering a concrete realization of holographic emergence where gravity arises from the quantum information structure of the boundary theory.
\end{abstract}
\maketitle

\tableofcontents

\section{Introduction}
The AdS/CFT correspondence, in its strongest form, states that ${\cal N}=4$, $D=3+1$ superconformal $SU(N)$ gauge theory is dual to type IIB string theory on the AdS$_5\times$S$^5$ background \cite{Maldacena:1997re, Witten:1998qj, Gubser:1998bc, Aharony:1999ti, DHoker:2002nbb, Casalderrey-Solana:2011dxg, Nastase:2007kj}. This correspondence establishes a remarkable duality between a theory of quantum gravity and a non-gravitational gauge theory providing a powerful framework for exploring the quantum nature of spacetime. At first glance, it presents a significant conceptual challenge: how can gravity emerge from a gauge theory when the numbers of degrees of freedom differ so drastically \cite{Takayanagi:2025ula}? Indeed, the gravitational theory has ${\cal O}(1)$ degrees of freedom corresponding to the graviton and other light fields, while the gauge theory possesses ${\cal O}(N^2)$ degrees of freedom associated with the color degrees of freedom. 
Moreover, in the 't Hooft limit where $N \to \infty$ and the 't Hooft coupling $\lambda = g_{\rm YM}^2 N$ is held large, the string theory reduces to classical type IIB supergravity which is dual to a strongly coupled gauge theory. This limit is particularly valuable because classical gravity is far more tractable than the strongly coupled field theory, enabling the computation of non-perturbative quantities that would be otherwise inaccessible. Consequently, holographic duality has been extensively applied to investigate the behavior of strongly coupled field theories, yielding profound insights into phenomena such as quark-gluon plasma, superconductivity and quantum criticality \cite{Casalderrey-Solana:2011dxg}.
Crucially, the question of how gravity emerges from the underlying gauge theory has been a central focus of recent research. A key quantity that provides a direct window into this emergence is holographic entanglement entropy (HEE) \cite{Headrick:2019eth, Calabrese:2004eu, Casini:2009sr, Ryu:2006bv, Hubeny:2007xt, Rangamani:2016dms, Nishioka:2009unn, Calabrese:2009qy, Nishioka:2009un} which encodes geometric information in the entanglement structure of the boundary field theory, as we shall discuss.

Entanglement entropy \cite{Calabrese:2009qy} quantifies the amount of quantum information shared between two subsystems, or equivalently, the information lost when tracing out one of them. Consider a pure state $\ket{\psi}$ partitioned into a subsystem $A$ and its complement $A^c$. The reduced density matrix is defined as $\rho_A = \operatorname{Tr}_{A^c}\left(\ket{\psi}\bra{\psi}\right)$ where the trace is performed over the degrees of freedom of the complementary subsystem. The von Neumann entanglement entropy is then given by $S_A = -\operatorname{Tr}\left(\rho_A \log \rho_A\right)$
which serves as a fundamental measure of the quantum correlations between the two subsystems. Entanglement entropy captures the non-local quantum correlations that are intrinsic to the quantum nature of the state. This quantity is central to quantum information theory and plays a crucial role in understanding the structure of quantum many-body systems, quantum phase transitions, and the thermodynamics of black holes. 

In the context of the AdS/CFT correspondence, these field-theoretic quantities usually admits a remarkably simple geometric dual. The HEE is computed by extremizing the area of a codimension-two surface \(\Gamma_A\) that extends into the bulk spacetime and whose boundary coincides with the entangling surface \(\partial A\) on the AdS boundary. The HEE is then given by the celebrated Ryu-Takayanagi (RT) formula \cite{Ryu:2006bv, Hubeny:2007xt}
\be\label{RT}
S_A = \frac{\operatorname{Area}(\Gamma_A)}{4G_N},
\ee
where \(\Gamma_A\) is the extremal surface anchored on \(\partial A\) and \(G_N\) is the Newton constant of the bulk gravitational theory. This elegant geometric prescription has proven to be a powerful tool for exploring the entanglement structure of strongly coupled quantum field theories and has been extensively applied to unravel various aspects of holographic duality. It establishes a profound connection between the information-theoretic concept of entanglement and the geometric properties of spacetime \cite{Takayanagi:2025ula}, reinforcing the idea that gravity may emerge from the underlying quantum entanglement structure.

Timelike entanglement entropy \cite{Doi:2023zaf} has been introduced and studied as a generalization of the standard spacelike entanglement entropy to scenarios involving temporal intervals \cite{Doi:2022iyj, Ali-Akbari:2026xzx, Chu:2025sjv,  Afrasiar:2024ldn, Goki:2026hpl}. Like its spacelike counterpart, this quantity measures the quantum entanglement between two subsystems, but now with the interval extending along the time direction rather than a spatial slice. In the context of the AdS/CFT correspondence, timelike entanglement was initially approached via an analytic continuation of the spatial HEE, which generically yields complex results interpreted as pseudoentropy \cite{Doi:2022iyj}. However, a more fundamental holographic prescription for timelike entanglement entropy (HTEE) was subsequently proposed \cite{Heller:2024whi}. This prescription closely parallels the RT formula \eqref{RT}, with the crucial distinction that the real extremal surface must be replaced by a complex one embedded in the complexified bulk geometry. Extremizing this complex surface yields both the real and imaginary parts of the HTEE.

In quantum information theory, the entanglement entropy counts the number of maximally entangled states, the Bell pairs. This provides a direct interpretation of the holographic principle where the maximum information content of a region is encoded by its boundary area \cite{Takayanagi:2025ula}. Since, in natural units, $G_N\approx \ell_P^2$ where $\ell_P$ is the Planck length, \eqref{RT} states how many Bell pairs exist in every Planck-scale area; or in other words, spacetime is filled with a finite number of Bell pairs per unit area. This suggests that spacetime is not a continuous manifold at the deepest level, but rather a discrete network of quantum entanglement bonds. Therefore the dynamics of spacetime can be related to the dynamics of entanglement entropy and the Einstein equations can describe the dynamics of entanglement entropy. In fact, varying the entanglement structure corresponds to perturbing the bulk geometry. Thus, gravity ceases to be a fundamental force and instead emerges as the macroscopic manifestation of quantum information processing. Therefore a fundamental question arises: how can we describe the emergence of spacetime using entanglement entropy? Several works have established a deep connection between the dynamics of entanglement entropy and the emergence of gravitational equations. In particular, it has been shown that the linearized Einstein equations can be understood as a basic dynamical consequence of entanglement entropy with the first law of entanglement entropy playing a central role \cite{Blanco:2013joa, Lashkari:2013koa, Faulkner:2013ica}. Beyond the linear order, this correspondence has been extended to the full nonlinear regime, where the Einstein equations emerge from the entanglement structure of the dual field theory \cite{Faulkner:2017tkh, Oh:2017pkr}. Furthermore, the foundational idea that spacetime geometry itself is built up from quantum entanglement has been articulated in \cite{VanRaamsdonk:2010pw}, providing a conceptual framework for understanding gravity as an emergent phenomenon. More recent developments have continued to explore and refine this connection \cite{Li:2025tud}, further solidifying the notion that gravity arises from the quantum information content of the underlying theory.

In this paper, we consider a thermal $1+1$-dimensional conformal field theory dual to an AdS$_3$ black hole, namely the BTZ black hole. We systematically compute the HEE and HTEE across all parameter regimes, covering both the exterior and interior of the black hole horizon. This yields four distinct entanglement measures, each probing a different region of the bulk geometry and capturing complementary aspects of the underlying spacetime.
Remarkably, by taking appropriate, first or second, derivatives of these entanglement measures with respect to the spacelike and timelike boundary intervals and the turning point of the extremal surface, we precisely reconstruct the various metric components both inside and outside the black hole. 
Our results provide a concrete realization of the inverse holographic problem, namely the reconstruction of the full gravitational spacetime from boundary entanglement entropy data. This work offers a significant step toward a deeper understanding of how gravity emerges from quantum information, as the Einstein equations arise naturally as a consequence of the holographic dictionary rather than being imposed from the outset.

\section{Holographic entanglement measures}
We start with an AdS$_3$ black hole metric
\be\label{metric}
ds^2 = -g_{tt}(z) \, dt^2 + \frac{dz^2}{g_{zz}(z)} + g_{xx}(z) \, dx^2,
\ee
where the metric components are given by
\bse\begin{align}
\label{gtt} g_{tt}(z) &= \frac{1}{z^2}\left(1 - \frac{z^2}{z_h^2}\right)=\frac{1}{z^2}f(z),\\
\label{gzz} g_{zz}(z) &= \frac{1}{z^2 \left(1 - \frac{z^2}{z_h^2}\right)}=\frac{1}{z^2 f(z)},\\
\label{gxx} g_{xx}(z) &= \frac{1}{z^2}.
\end{align}\ese
The radial direction is denoted by $z$ and the dual field theory lives on the 1+1-dimensional boundary of this geometry located at $z=0$. The black hole horizon is situated at $z = z_h$ and the Hawking temperature, which is identified with the temperature of the dual field theory, is given by
\be
T = \frac{1}{2\pi z_h}.
\ee
In the following, we shall utilize this metric to compute various entanglement measures in the field theory, considering both spacelike and timelike intervals. These measures will subsequently serve as the input data for the metric reconstruction procedure.

In order to compute the HEE for a single interval, we employ the RT conjecture, which equates the entanglement entropy of a boundary region to the area of the extremal codimension-two surface anchored on its boundary. We consider the following spacelike interval on the boundary time slice,
\be\label{intervalx}
A_s : \left\{ x \in \left[-\frac{\Delta x}{2}, \frac{\Delta x}{2}\right], \ t = 0 \right\}.
\ee
Given the static nature of the background, the extremal surface can be parametrized by the radial coordinate $z$. Consequently, the profile $x(z)$ completely describes the bulk surface. Substituting the metric components \eqref{metric} into the proper length functional, the geodesic length to be extremized is given by
\be
L = \int \frac{1}{z} \sqrt{\frac{1}{f(z)} + \left( \frac{dx}{dz} \right)^2 } \, dz.
\ee
Since the coordinate $x$ does not appear explicitly in the integrand, it is a cyclic variable, leading to a conserved conjugate momentum
\be
P = \frac{\partial L}{\partial \dot{x}}.
\ee
Solving this equation for $\dot{x} \equiv dx/dz$ yields
\be \label{dot}
\dot{x} = \frac{P z}{\sqrt{f(z)\left(1 - P^2 z^2\right)}}.
\ee
We define the turning point $z_*$ as the maximum radial depth reached by the geodesic, where $\dot{x} \to \infty$. This immediately gives
\be\label{zstars} 
z_* = \frac{1}{P}.
\ee
This relation shows that the turning point $z_*$ is independent of the black hole horizon $z_h$. This turning point marks the deepest penetration of the geodesic into the bulk and is the point at which the surface smoothly bounces back toward the boundary. Imposing the boundary conditions $x(\epsilon) = \pm \Delta x / 2$, where $\epsilon$ is a UV cutoff near the AdS boundary, we integrate $dx = \dot{x} \, dz$ from the cutoff up to the turning point. The total boundary separation is thus expressed as an integral that relates the interval length to the radial depth of the geodesic. Substituting $f(z) = 1 - z^2/z_h^2$, we obtain the standard integral
\be\label{HEEf}
\frac{\Delta x}{2} = \int_{\epsilon}^{z_*} \frac{Pz}{ \sqrt{\left(1 - \frac{z^2}{z_h^2}\right)\left(1 - P^2z^2\right)}} \, dz=\frac{z_h}{2}\ln\left|\frac{Pz_h-1}{Pz_h+1}\right|.
\ee
This integral encodes the scale-radius duality of holography: smaller boundary intervals probe near-boundary regions (small $z_*$), while larger intervals penetrate deeper into the bulk (larger $z_*$). In the following, we analyze this integral for two different regimes, namely, the exterior of the horizon ($z_* < z_h$) and the interior of the horizon ($z_* > z_h$) and explain their physical properties separately, as these correspond to the standard HEE and its complementary spacelike counterpart, respectively.

\subsection{Holographic spacelike entanglement entropy}
Since the absolute value appears in \eqref{HEEf}, we must treat the two regimes $Pz_h>1$ and $Pz_h<1$ separately. These two cases correspond geometrically to the exterior and interior of the black hole horizon, respectively, as we shall demonstrate below.

\begin{itemize}
\item {{\it Solution with $Pz_h>1\ (z_h>z_*)$}}:\\
Focusing first on the exterior regime, we evaluate the integral in \eqref{HEEf} without the absolute value, since the argument of the logarithm is positive. Carrying out the integration yields a logarithmic expression that directly relates the boundary interval $\Delta x$ to the conserved momentum $P$
\be
\Delta x = z_h \ln\left[\frac{P z_h - 1}{P z_h + 1}\right].
\ee
Inverting this logarithmic relation using the definition of the turning point $P = 1/z_*$, we immediately obtain the standard duality
\be\label{zstars_ext}
z_* = z_h \tanh\left(\frac{\Delta x}{2z_h}\right).
\ee
This explicit form of $z_*$ reveals how the boundary interval length controls the penetration depth of the extremal surface into the bulk. Substituting the geodesic profile \eqref{dot} into the length functional and subsequently applying the RT formula \eqref{RT} together with the turning point relation \eqref{zstars_ext}, we obtain the HEE for the spacelike interval
\be\label{HEE0}
S^{\rm HEE}_{\Delta x} = \frac{c}{3} \ln\left[\frac{2z_h}{\epsilon} \sinh\left(\frac{\Delta x}{2z_h}\right)\right].
\ee
Here, the parameter $c$ denotes the central charge of the dual CFT$_2$ and is identified as $c = \frac{3}{2G_N}$. This solution therefore corresponds to the standard HEE for a spacelike interval in the exterior of the black hole.

\item {\it Solution with $Pz_h<1\ (z_h<z_*)$}:\\
Following the same procedure as in the previous case, but now taking the absolute value in the opposite regime \(Pz_h < 1\), we obtain from \eqref{HEEf} the modified relation between the boundary interval and the conserved momentum
\be
\Delta x = z_h \ln\left[\frac{1 - P z_h}{1 + P z_h}\right].
\ee
Notably, this expression yields a positive boundary separation for \(Pz_h < 1\), as both the numerator and denominator are positive. Inverting this logarithmic relation using \(P = 1/z_*\) gives the corresponding turning point
\be\label{zss}
z_* = z_h \coth\left(\frac{\Delta x}{2z_h}\right),
\ee
which is manifestly greater than the horizon radius. This indicates that the extremal surface now probes the region behind the black hole horizon, corresponding to the \textit{interior} of the geometry. This complementary solution describes a distinct spacelike geodesic that connects points on the boundary while passing through the interior. Substituting this profile into the length functional and applying the RT formula yields the complementary HEE$^s$
\be\label{HEEss}
S^{{\rm HEE}^s}_{\Delta x} = \frac{c}{3} \ln\left[\frac{2z_h}{\epsilon} \cosh\left(\frac{\Delta x}{2z_h}\right)\right].
\ee
Remarkably, this quantity remains real for all values of \(\Delta x\). The appearance of the hyperbolic cosine in the prefactor reflects the change in the global structure of the geodesic as it extends into the interior, encoding information about the deep bulk physics. Together, the exterior and interior solutions provide a complete holographic description of the spacelike entanglement structure across the entire geometry.

\end{itemize}

\subsection{Holographic timelike entanglement entropy}
We now turn our attention to the computation of the HTEE. To this end, we consider a timelike interval along the boundary time direction, defined as
\be\label{intervalt}
A_t : \left\{ t \in \left[-\frac{\Delta t}{2}, \frac{\Delta t}{2}\right],\ x = 0 \right\}.
\ee
Unlike the spacelike case, this interval extends along the time direction and its holographic dual requires a complexified bulk geometry. Using this configuration, the length functional to be extremized is given by
\be 
L = \int \frac{dz}{z} \sqrt{-f(z) \, \dot{t}^2 + \frac{1}{f(z)}},
\ee
where the minus sign under the square root reflects the timelike nature of the boundary interval and necessitates a careful treatment of the square root branch. Since the coordinate $t$ does not appear explicitly in the integrand, it is a cyclic variable, leading to a conserved conjugate momentum
\be 
E = \frac{\partial L}{\partial \dot{t}}.
\ee
Solving this equation for $\dot{t} \equiv dt/dz$ yields
\be
\dot{t} = \frac{z E}{f(z) \sqrt{f(z) + z^2 E^2}}.
\ee
The turning point of the extremal surface is determined by the condition $\dot{t} \to \infty$ which requires the denominator to vanish, giving the algebraic equation
\be
f(z_*) + z_*^2 E^2 = 0.
\ee
Crucially, this condition implies that the turning point $z_*$ is generally complex, as $f(z) = 1 - z^2/z_h^2$ and $E$ are real parameters. This complexification is a hallmark of timelike entanglement. Imposing the boundary conditions $t(\epsilon) = \pm \Delta t / 2$, we integrate $dt = \dot{t} \, dz$ from the cutoff to the turning point. The total boundary separation is thus expressed as
\be\label{lt} 
\frac{\Delta t}{2} = \int_\epsilon^{z_*}\frac{z E}{\left(1-\frac{z^2}{z_h^2}\right)\sqrt{1+\left(E^2-\frac{1}{z_h^2}\right)z^2}} \, dz = \frac{z_h}{2}\ln\left|\frac{E z_h + 1}{E z_h - 1}\right|.
\ee
This integral relation directly links the timelike boundary interval to the conserved momentum. Depending on the magnitude of $E z_h$, the turning point $z_*$ may lie either in the exterior or in the interior of the horizon or become purely imaginary. In the following, we analyze these different regimes to extract the corresponding timelike entanglement entropies.

\begin{itemize}
\item {{\it Solution with $Ez_h>1\ (|z_*|<z_h)$}}:\\
Using the integral relation obtained above, we can explicitly relate the timelike boundary interval $\Delta t$ to the conserved momentum $E$. Evaluating the logarithmic expression gives
\be
\Delta t = z_h \ln\left[\frac{E z_h + 1}{E z_h - 1}\right].
\ee
As in the spacelike case, the absolute value in the logarithmic argument requires us to consider two distinct regimes depending on the magnitude of $E z_h$. Inverting this relation using the turning point condition $f(z_*) + z_*^2 E^2 = 0$, we find that the turning point becomes purely imaginary
\be\label{zHTEE}
z_* = i z_h \sinh\left(\frac{\Delta t}{2z_h}\right).
\ee
This complexification of the turning point is a direct consequence of the timelike nature of the boundary interval and signals that the extremal surface now extends into the complexified bulk geometry. Substituting this complex turning point into the length functional and applying the RT formula yields the HTEE
\be
S^{\rm HTEE}_{\Delta t} = \frac{c}{3} \ln\left[\frac{2z_h}{\epsilon} \sinh\left(\frac{\Delta t}{2z_h}\right)\right] + \frac{i\pi c}{6}.
\ee
Remarkably, the entropy acquires an imaginary part, $i\pi c/6$, which serves as a universal hallmark of timelike entanglement in holographic two-dimensional conformal field theories. Importantly, this imaginary contribution emerges directly from the complex geometry of the timelike extremal surface whose turning point is intrinsically complex, as given in \eqref{zHTEE}. 
\item {{\it Solution with $Ez_h<1\ (z_h<z_*)$}}:\\
In this regime, the absolute value in the logarithmic expression \eqref{lt} must be evaluated with opposite sign, yielding
\be 
\Delta t = z_h \ln\left[\frac{1 + E z_h}{1 - E z_h}\right].
\ee
Inverting this relation using the turning point condition \(f(z_*) + z_*^2 E^2 = 0\), we obtain the corresponding turning point
\be
z_* = z_h \cosh\left(\frac{\Delta t}{2z_h}\right).
\ee
Since the hyperbolic cosine satisfies \(\cosh(\Delta t / 2z_h) \geq 1\) for all real \(\Delta t\), we have \(z_* \geq z_h\), with equality only in the limit \(\Delta t \to 0\). This indicates that the extremal surface now extends into the interior of the black hole horizon, probing the region \(z > z_h\). This is the timelike counterpart of the complementary spacelike solution discussed earlier. Substituting the corresponding profile into the length functional and applying the RT formula yields the HEE$^t$ for this timelike interval:
\be
S^{{\rm HEE}^t}_{\Delta t} = \frac{c}{3} \ln\left[\frac{2z_h}{\epsilon} \cosh\left(\frac{\Delta t}{2z_h}\right)\right].
\ee
Notably, this entropy remains completely real for all values of \(\Delta t\), in contrast to the previous timelike solution which acquired an imaginary part. This real-valued nature reflects the fact that the extremal surface lies entirely within the real geometry, even though it passes through the interior of the horizon. Together with the complex exterior solution, this complementary timelike measure provides a complete holographic description of the timelike entanglement structure across the entire black hole geometry, encompassing both the exterior and interior regions of the black hole.
\end{itemize}

\section{Emergence of bulk geometry}
So far, we have computed four different types of holographic entanglement measures, namely, HEE, HEE$^{s}$, HTEE and HEE$^{t}$, in a thermal 1+1-dimensional conformal field theory. In fact, by utilizing the RT conjecture, we analytically derived the entanglement entropy for various spacelike and timelike intervals, as well as their complements, within the AdS$_3$ black hole background, covering both the exterior and interior of the event horizon. From now on, we set aside the gravitational dual and aim to reconstruct the metric components of the dual gravity background using only the entanglement data obtained from the field theory. In other words, we demonstrate that the collected entanglement information is sufficient to uniquely determine the three components of the background metric. To proceed, let us first summarize our entanglement entropy results:
\bse\begin{align}
\label{HEE} S^{\rm HEE}_{\Delta x} &= \frac{c}{3} \ln\left[\frac{2z_h}{\epsilon} \sinh\left(\frac{\Delta x}{2z_h}\right)\right],\\
\label{HEEx} S^{{\rm HEE}^s}_{\Delta x} &= \frac{c}{3} \ln\left[\frac{2z_h}{\epsilon} \cosh\left(\frac{\Delta x}{2z_h}\right)\right],\\
\label{HTEE} S^{\rm HTEE}_{\Delta t} &= \frac{c}{3} \ln\left[\frac{2z_h}{\epsilon} \sinh\left(\frac{\Delta t}{2z_h}\right)\right] + \frac{i\pi c}{6},\\
\label{HEEt} S^{{\rm HEE}^t}_{\Delta t} &= \frac{c}{3} \ln\left[\frac{2z_h}{\epsilon} \cosh\left(\frac{\Delta t}{2z_h}\right)\right].
\end{align}\ese
Apart from the third one, the other three entropies yield real-valued solutions. Notably, both \eqref{HTEE} and \eqref{HEEt} are obtained in the complexified geometry, indicating that timelike entanglement probes regions that are not accessible to purely spacelike definitions. Furthermore, it is important to note that choosing specific values for the spacelike interval $\Delta x$ and timelike interval $\Delta t$ corresponds to distinct turning points for each case, given explicitly by
\bse\begin{align}
\label{star1} z_* &=z_h\tanh\left(\frac{\Delta x}{2z_h}\right),\\
\label{star2} z_* &=z_h\coth\left(\frac{\Delta x}{2z_h}\right),\\
\label{star3} z_* &=iz_h\sinh\left(\frac{\Delta t}{2z_h}\right),\\
\label{star4} z_* &=z_h\cosh\left(\frac{\Delta t}{2z_h}\right).
\end{align}\ese
These equations explicitly show that deeper points in the bulk geometry are probed by larger values of the time and space intervals. In particular, equations \eqref{star2} and \eqref{star4} indicate that regions behind the horizon become accessible through the use of HTE$^s$ and HEE$^{t}$. Consequently, every point in the bulk can be systematically reached by appropriately choosing $\Delta t$ and $\Delta x$, or equivalently, by selecting the corresponding entanglement measures. This underscores the fact that the complete set of entanglement entropies forms a holographic ``dataset'' capable of reconstructing the entire bulk spacetime, thereby providing a direct path from quantum information to emergent geometry.

What we aim to do now is to take the second derivatives of the entanglement measures we have obtained. Since we are dealing with the simplest 1+1-dimensional conformal field theory, which depends only on the coordinates $t$ and $x$, the extraction of metric components proceeds via a straightforward derivative prescription. In particular, the second derivative of the entanglement entropy with respect to the spacelike interval $\Delta x$ yields the $tt$-component of the bulk metric, while the second derivative with respect to the timelike interval $\Delta t$ yields the $xx$-component. This cross-identification originates from the fact that the extremal surface probes the transverse directions, allowing the entanglement data to encode the local geometry. Therefore, our metric components are recovered in this elegant manner. To see this explicitly, let us start with the HEE. Taking the second derivative of \eqref{HEE} and substituting the corresponding turning point from \eqref{star1} into the result, we obtain
\be
g_{tt*}=\frac{12}{c}\frac{\partial^2 S^{\rm HEE}_{\Delta x}}{\partial(\Delta x)^2} = 
-\frac{1}{z_*^2}\left(1-\frac{z_*^2}{z_h^2}\right)<0,\ \ z_*>z_h,
\ee
where $g_{tt*}=g_{tt}(z=z_*)$ and so on. The coefficient $12/c$ is chosen such that the correct metric component is produced without any extra constant factor. As is clearly seen, this corresponds precisely to the metric component \eqref{gtt} evaluated at $z=z_*$ for the exterior of the horizon. Similarly, starting with \eqref{HEEx} and using \eqref{star2}, we get
\be
g_{tt*}=\frac{12}{c}\frac{\partial^2 S^{{\rm HEE}^s}_{\Delta x}}{\partial(\Delta x)^2} = 
-\frac{1}{z_*^2}\left(1-\frac{z_*^2}{z_h^2}\right)>0,\ \ z_*<z_h,
\ee
which corresponds to the interior of the black hole horizon. In summary, we have
\be
g_{tt*}=
\begin{cases}
\frac{12}{c}\frac{\partial^2 S^{{\rm HEE}}_{\Delta x}}{\partial(\Delta x)^2} = -\frac{1}{z_*^2}\left(1-\frac{z_*^2}{z_h^2}\right)<0,\ \ z_*>z_h \\[6pt]
\frac{12}{c}\frac{\partial^2 S^{{\rm HEE}^s}_{\Delta x}}{\partial(\Delta x)^2}=-\frac{1}{z_*^2}\left(1-\frac{z_*^2}{z_h^2}\right)>0,\ \ z_*<z_h.
\end{cases}
\ee
Therefore, the information contained in our entanglement measures completely specifies the time component of the bulk metric for both sides of the horizon. In the same way, using the set \eqref{HTEE} and \eqref{star3} for the timelike entanglement and the set \eqref{HEEt} and \eqref{star4} for its counterpart, we find
\be
g_{xx*}=
\begin{cases}
\frac{12}{c}\frac{\partial^2 S^{\rm HTEE}_{\Delta t}}{\partial(\Delta t)^2} = \frac{1}{z_*^2},\ \ z_*>z_h \\[6pt]
\frac{12}{c}\frac{\partial^2 S^{{\rm HEE}^t}_{\Delta t}}{\partial(\Delta t)^2} = \frac{1}{z_*^2},\ \ z_*<z_h.
\end{cases}
\ee
which yield the spatial metric component \eqref{gxx} evaluated at $z=z_*$. Remarkably, this demonstrates that the full BTZ black hole geometry, including both the exterior and interior regions, can be systematically reconstructed solely from the second derivatives of the various entanglement entropies computed in the dual CFT. This provides a concrete realization of holographic emergence where the gravitational background is explicitly derived from the underlying quantum information structure, without any prior assumption about the bulk geometry.

To provide a clear overview of the metric reconstruction procedure, we summarize our findings in the table \ref{tabel}. This table systematically organizes the various types of entanglement measures, the nature of their associated intervals, the corresponding turning points and the regions of the black hole geometry they probe.
\begin{table}[htbp]
\centering
\caption{Reconstruction of the interior and exterior metric components of the black hole from entanglement measures}
\label{tabel}
\begin{tabular}{|c|c|c|c|c|c|c|}
\hline
{\it Entanglement} & {\it Interval} & {\it Condition} & {\it Real/Complex} & {\it Region Probed} & {\it Derivative w. r. t.} & {\it Emergent metric component}\\
\hline
HEE & Spacelike & $Pz_h>1$ & Real & Exterior & $\Delta x$ & $-g_{tt}<0$ \\
\hline
HEE$^s$ & Spacelike & $Pz_h<1$ & Real & Interior & $\Delta x$  & $-g_{tt}>0$ \\
\hline
HTEE & Timelike & $Ez_h>1$ & Complex & Exterior & $\Delta t$ & $g_{xx}>0$\\
\hline
HEE$^t$ & Timelike & $Ez_h<1$ & Real & Interior & $\Delta t$ & $g_{xx}>0$ \\
\hline
HEE & Spacelike & $Pz_h>1$ & Real & Exterior & $z_*$ & $g_{zz}>0$ \\
\hline
HEE$^s$ & Spacelike & $Pz_h<1$ & Real & Interior & $z_*$ & $g_{zz}<0$ \\
\hline
\end{tabular}
\end{table}
As the table clearly demonstrates, the exterior region is reconstructed from the real, spacelike HEE, while the interior region requires the use of the complementary spacelike measure HEE$^s$. Similarly, the timelike measures HTEE and HEE$^t$ distinguish between the exterior and interior through their complex and real nature, respectively. This unified picture confirms that the full black hole geometry, including its horizon, is encoded in the entanglement structure of the dual field theory. Notably, the conditions $Pz_h>1$ ($Ez_h>1$) and $Pz_h<1$ ($Ez_h<1$) correspond to the turning points lying outside or inside the horizon, thereby providing a direct diagnostic for probing the deep bulk.

The radial metric component $g_{zz}$ cannot be determined solely from the local information encoded in the second derivatives of the entanglement measures with respect to the boundary intervals $\Delta x$ and $\Delta t$. Unlike $g_{tt}$ and $g_{xx}$, which are fixed by the local curvature of the extremal surface at its turning point, the reconstruction of $g_{zz}$ requires a different approach. However, contrary to the naive expectation, no additional assumptions or external input are needed. The radial component can be extracted directly from the entanglement entropy itself, provided we take the derivative with respect to the correct variable: the turning point $z_*$, rather than the boundary interval. Therefore, the key to obtaining $g_{zz}$ lies in differentiating the on-shell entropy with respect to $z_*$, which effectively probes the radial direction of the bulk geometry.

To extract $g_{zz}$ from the entanglement data, we start with the exterior turning-point relation \eqref{zstars_ext} and substitute it into the spacelike HEE \eqref{HEE0}. Using the identity $\sinh(\operatorname{arctanh}(y)) = y/\sqrt{1-y^2}$ with $y = z_*/z_h$, we obtain
\be
S^{\rm HEE}_{\Delta x}(z_*) = \frac{c}{3} \ln\left[\frac{2z_*}{\epsilon\sqrt{1 - z_*^2/z_h^2}}\right], \qquad z_* < z_h.
\ee
Similarly, for the complementary spacelike solution using \eqref{zss} and \eqref{HEEss}, the identity $\cosh(\operatorname{arccoth}(y)) = y/\sqrt{y^2 - 1}$ yields
\be
S^{{\rm HEE}^s}_{\Delta x}(z_*) = \frac{c}{3} \ln\left[\frac{2z_*}{\epsilon\sqrt{z_*^2/z_h^2 - 1}}\right], \qquad z_* > z_h.
\ee
Differentiating these expressions with respect to $z_*$, we obtain
\be
g_{zz}(z_*) =
\begin{cases}
\displaystyle \frac{3}{c z_*}\frac{\partial S^{\rm HEE}_{\Delta x}}{\partial z_*} = \frac{1}{z_*^2\left(1 - \frac{z_*^2}{z_h^2}\right)} > 0, & z_* < z_h, \\[12pt]
\displaystyle \frac{3}{c z_*}\frac{\partial S^{{\rm HEE}^s}_{\Delta x}}{\partial z_*} = \frac{1}{z_*^2\left(1 - \frac{z_*^2}{z_h^2}\right)} < 0, & z_* > z_h.
\end{cases}
\ee
Remarkably, this reproduces the radial metric component \eqref{gzz} exactly, without any additional assumptions or external input. The key observation is that the derivative must be taken with respect to the turning point $z_*$, the variable that encodes the depth of the extremal surface in the bulk, rather than the boundary interval. This procedure relies solely on the HEE and the RT formula, demonstrating that the full bulk geometry, including its radial structure, is encoded in the entanglement data of the dual field theory.

\begin{table}[htbp]
\centering
\caption{Reconstruction of the metric components from entanglement measures in the zero-temperature limit \(z_h \to \infty\).}
\label{tab:zero_temp_reconstruction}
\begin{tabular}{|c|c|c|c|c|c|c|}
\hline
{\it Entanglement} & {\it Interval} & {\it Condition} & {\it Real/Complex} & {\it Region Probed} & {\it Derivative w.r.t.} & {\it Emergent metric component} \\
\hline
HEE & Spacelike & \(z_* = \frac{\Delta x}{2}\) & Real & Exterior & \(\Delta x\) & \(-g_{tt} <0 \) \\
\hline
HEE\(^s\) & Spacelike & \(z_* \to \infty\) & Real & \multicolumn{1}{c|}{---} & \multicolumn{1}{c|}{---} & \multicolumn{1}{c|}{---} \\
\hline
HTEE & Timelike & \(z_* = i\frac{\Delta t}{2}\) & Complex & Exterior & \(\Delta t\) & \(g_{xx} > 0\) \\
\hline
HEE\(^t\) & Timelike & \(z_* \to \infty\) & Real & \multicolumn{1}{c|}{---} & \multicolumn{1}{c|}{---} & \multicolumn{1}{c|}{---} \\
\hline
HEE & Spacelike & \(z_* = \frac{\Delta x}{2}\) & Real & Exterior & \(z_*\) & \(g_{zz} > 0\) \\
\hline
HEE\(^s\) & Spacelike & \(z_* \to \infty\) & Real & \multicolumn{1}{c|}{---} & \multicolumn{1}{c|}{---} & \multicolumn{1}{c|}{---} \\
\hline
\end{tabular}
\end{table}

It is instructive to examine the zero-temperature limit \(z_h \to \infty\) of our results. In this limit, the BTZ black hole geometry reduces to pure AdS\(_3\) spacetime. Our metric components \eqref{gtt}, \eqref{gxx}, and \eqref{gzz} consistently reduce to \(g_{tt} = g_{xx} = g_{zz} = 1/z^2\), reproducing the well-known Poincar\'e metric. The standard HEE \eqref{HEE} reduces to the familiar vacuum CFT\(_2\) result \(S_{\Delta x} = \frac{c}{3}\ln(\Delta x/\epsilon)\), while the timelike HTEE \eqref{HTEE} yields the analytically continued expression \(\frac{c}{3}\ln(\Delta t/\epsilon) + i\pi c/6\), which is the pseudoentropy for a timelike interval in the vacuum. The complementary spacelike and timelike entropies \eqref{HEEx} and \eqref{HEEt} diverge logarithmically, reflecting the fact that the complementary geodesics probing the interior of the black hole become infinitely long when the horizon recedes to infinity. Similarly, the complementary turning points \eqref{star2} and the timelike counterpart diverge in this limit. These behaviors are entirely consistent with the expected pure AdS geometry, thereby providing a stringent consistency check of our holographic reconstruction procedure. Table \ref{tab:zero_temp_reconstruction} summarizes the finite metric components reconstructed from the surviving entanglement measures in this limit.

At the end of this paper, let us review two important points.

\begin{itemize}
\item In order to obtain the radial component of the metric, $g_{zz}$, it is natural and physically reasonable to take the derivative of the entanglement measures with respect to the radial coordinate $z_*$. This is because the derivative with respect to $z_*$ directly probes variations along the holographic direction and this variation precisely yields the radial metric component. In other words, the turning point $z_*$ encodes the depth of the extremal surface in the bulk and differentiating the entropy with respect to this depth extracts the geometry of the radial direction.

\item As we have demonstrated, the second derivative of the entanglement measures with respect to a spacelike interval $\Delta x$ yields the timelike metric component $g_{tt}$, while the derivative with respect to a timelike interval $\Delta t$ yields the spacelike component $g_{xx}$. This cross-identification may appear counterintuitive at first glance, as one might naively expect the spacelike (timelike) entanglement measures to give the spacelike (timelike) component of the metric. However, we emphasize that this is a direct consequence of the mathematical structure of holographic duality and the specific form of the BTZ geometry. We do not claim a deeper physical principle beyond what the mathematics dictates.
\end{itemize}

\section*{Acknowledgments}
We would like to thank DeepSeek for its assistance in improving the presentation of this manuscript.

\end{document}